\documentclass[10pt,showpacs,amsmath,amssymb,floatfix,superscriptaddress,longbibliography,twocolumn,eqsecnum]{revtex4-2}
\usepackage{amsmath}
\usepackage{physics}
\usepackage{graphicx}
\usepackage[usenames,dvipsnames,svgnames,table]{xcolor}
\usepackage{tcolorbox}
\usepackage{booktabs}
\usepackage{makecell} 
\usepackage{tabularx}
\usepackage{chngcntr}
\usepackage[utf8]{inputenc}
\counterwithout{equation}{section}
\counterwithout{figure}{section}
\usepackage[unicode=true,
            pdfusetitle,
            bookmarks=true,
            bookmarksnumbered=false,
            bookmarksopen=false,
            breaklinks=true,
            pdfborder={0 0 0},
            backref=false,
            colorlinks=true,
            hypertexnames=false]{hyperref}
\hypersetup{linkcolor=NavyBlue,urlcolor=NavyBlue,citecolor=NavyBlue}

\begin{document}
	
\title{The non-equilibrium condensate-like state in multi-mode driven dissipative superconducting quantum circuit}
	
\author{Jing Li}
\affiliation{College of Physics, Hangzhou Dianzi University, Hangzhou 310018, China}
\author{Chaoying Zhao}
\email{zchy49@163.com}
\affiliation{College of Physics, Hangzhou Dianzi University, Hangzhou 310018, China}
\affiliation{State Key Laboratory of Quantum Optics Technologies and Devices, Shanxi University, Taiyuan, 030006,China}
\affiliation{ Zhejiang Key Laboratory of Quantum State Control and Optical Field Manipulation, Hangzhou Dianzi University, Hangzhou, 310018, China}
\date{\today}

\begin{abstract}
The effective decay rate of traditional single-mode model for an open quantum system can't accurately described the exchange between inter-modes and dissipation caused by the environment, therefore we can not clearly see the occupation pathway of target mode or the transient participation of auxiliary modes. At the same time, we also can't address the selective dissipation pathways resulting from the circuit design. In order to solving above problem, we adopt fluxonium-transmon-transmon (FTT)-based three-mode model to describe dynamical characteristics of dissipation spectrum in superconducting quantum circuits. In terms of periodic flux modulation, we find out the pathways of inter-mode occupation transfer and environmental release for the target fluxonium changes from a high-occupation non-equilibrium condensate-like state to low-occupation equilibrium Bose–Einstein condensation(BEC) state. Without increasing the intrinsic loss of the target mode in the process of adjusting the auxiliary mode, we furthermore obtain additional selectively engaged environmental dissipative pathway. Our model can provides a new design perspective for non-equilibrium state recovery in a controllable dissipation superconducting quantum circuit. 
\end{abstract}

\maketitle

\section{INTRODUCTION}
 Bose–Einstein condensation(BEC) is a equilibrium phenomenon, which is characterized by occupation of bosonic modes. The equilibrium states determined by thermodynamic, depend on the balance between energy injection and environmental loss, allowing dissipation to participate directly in state formation and stabilization \cite{Petiziol2022}. At the same time, the competition between driving, nonlinearity, and dissipation result in the non-equilibrium states. Driven open quantum systems can explore non-equilibrium states. Experimentally, Josephson nonlinearity combined with multi-photon driving can support stable high occupation states in a single mode \cite{Hajr2024}. These states are not equilibrium BEC states due to  excitation number is not conserved and continuous driving is required, we call them condensate-like states. Although the Kerr model or parametric-oscillator model has introduced an effective decay rate result from environmental effects can encapsulate the threshold dynamics, population expansion, and equilibrium saturation. Nevertheless, it is unable to address the selective dissipation pathways result from the circuit design, and monitor the transfer of surplus occupation across various modes. 
 
  Superconducting quantum circuits offer a controllable platform for investigating coherent exchange, spectral control, and environmental dissipation in a multi-modes. Recent fluxonium-transmon coupler experiments show that inter-mode exchange can be tuned while preserve the long coherence of the target mode \cite{Ding2023}. Reservoir-engineering studies further show that controlled loss through auxiliary degrees of freedom can stabilize entangled states and Floquet-engineered bosonic dynamics \cite{Brown2022}. 
  Although multi-mode open-system models can capture the participation of auxiliary degrees of freedom, how does a target non-equilibrium state evolve when an auxiliary dissipative connection is introduced only temporarily, and can it re-approach its original non-equilibrium state after part of the target-mode occupation has already been irreversibly released to the environment? This process is naturally described in a multi-mode framework that retains both coherent inter-mode exchange and mode-resolved environmental release.
  
  In this paper, we investigate a three-mode superconducting circuit composed of a target fluxonium, a flux-tunable intermediate transmon, and a dissipative transmon. The target fluxonium supports a high-occupation non-equilibrium state established by the combined action of Kerr nonlinearity, an effective two-photon driving, and dissipation. Flux tuning of the intermediate transmon controls the spectral connection to the auxiliary dissipative pathway, while the dissipative transmon releases transfer occupation into environment. Periodic modulation temporarily enhances the auxiliary dissipative pathway and drives the target mode into a low-occupation region. After this pathway is spectrally suppressed, the original balance among two-photon driving, Kerr nonlinear dynamics, and weak auxiliary loss is restored, allowing the target mode to re-approach its original high-occupation state. This recovery does not reverse the environmental release, it proves recover-ability of the original driven state after a transient dissipative excursion. We use three-mode Lindblad dynamics to relate target-state evolution and mode-resolved transport, to examine the robustness of depletion and recovery against pure dephasing, finite temperature, and detuning fluctuations.

\section{THEORETICAL MODEL}
 Recently, fluxonium-transmon \cite{Manucharyan2009,Koch2007} tunable-coupler implementations motivate the use of an intermediate mode for spectral control \cite{Moskalenko2022}. We use a controllable auxiliary channel is related to reservoir-engineering and tunable-coupler approaches \cite{Petiziol2022,Moskalenko2022}. Our circuit are based on the fluxonium-transmon-transmon (FTT) architecture \cite{ZhangZhao2025}, in which the three modes are composed of a target fluxonium mode $F$, a flux-tunable intermediate transmon mode $T_1$, and a dissipative transmon mode $T_2$. The $F$ mode encompasses the nonlinear parametric dynamics, the $T_1$ mode regulates the intensity coupling between the $F$ mode and the supplementary dissipation pathway via frequency adjustment. The $T_2$ mode as a highly dissipative endpoint to expel surplus excitations into the surrounding environment.  As illustrated in Fig.1(a), the left node $1$ consists of a shunt capacitance $C_1$, a linear inductance $L_1$, and a Josephson junction with energy $E_{J_1}$, and forms the target fluxion \cite{Nguyen2019}. The central node $c$ contains a shunt capacitance $C_c$ and a direct current(DC) superconducting quantum interference device (SQUID), in which low-energy defines the flux-tunable intermediate transmon mode $T_1$. The right node $2$ contains a shunt capacitance $C_2$ and a Josephson junction with energy $E_{J_2}$, and forms the dissipative transmon mode $T_2$ through strong coupling with the surrounding electromagnetic environment. 

\begin{figure}[htbp]
	\centering
	\includegraphics[width=0.8\linewidth]{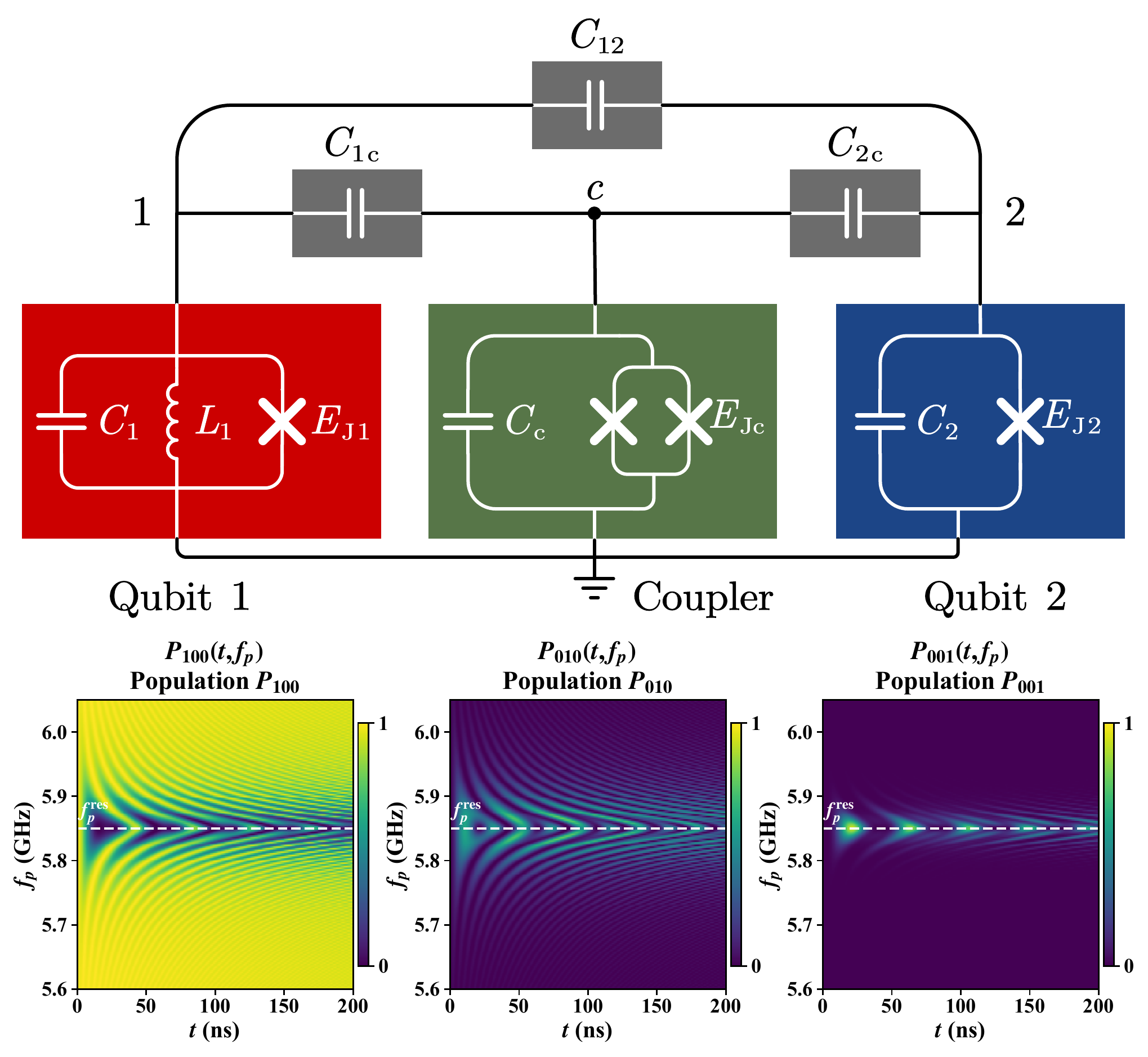}
	\caption{\label{fig:ftt_overview}The three-mode circuit architecture and single-excitation population transfer. (a) The circuit consists of a target fluxonium, a flux-tunable intermediate transmon, and a dissipative transmon. The capacitances $C_{1c}$, $C_{2c}$, and $C_{12}$ describe the capacitive couplings between the corresponding circuit nodes, rspectively. (b)-(d) Time-frequency maps of the single-excitation populations $P_{100}$, $P_{010}$, and $P_{001}$, respectively, which shows a redistribution of excitation among the target fluxonium, intermediate transmon, and dissipative transmon near the modulation resonance.}
\end{figure}

Under the rotating-wave approximation, we use the reduced node fluxes $\hat{\varphi}_j$ and Cooper-pair number operators $\hat{n}_j$ to describe the Hamiltonian of our system 
\begin{equation}
	\begin{split}
		\hat{H}_{cir} &= 4E_{C,F}\hat{n}_{1}^{2}+\frac{E_{L,F}}{2}\left( \hat{\varphi}_1-\hat{\varphi}_{F} \right) ^2-E_{J,F}\cos\hat{\varphi}_1 \\
		&+4E_{C,1}\hat{n}_{c}^{2}-E_{J_1}\varPhi _{c} \cos\hat{\varphi}_c \\
		&+4E_{C,2}\hat{n}_{2}^{2}-E_{J_2}\cos\hat{\varphi}_2 \\
&+g_{1c}\hat{n}_1\hat{n}_c+g_{2c}\hat{n}_2\hat{n}_c+g_{12}\hat{n}_1\hat{n}_2,
	\end{split}
	\label{eq:hamiltonian}
\end{equation}
Here, $E_{C,F}$, $E_{C,1}$, and $E_{C,2}$ are the charging energies of three nodes, $E_{L,F}$ is the inductive energy of the fluxonium, and $E_{J,F}$, $E_{J_1}\varPhi _{c}$, and $E_{J_2}$ are the corresponding Josephson energies, respectively. The charge-coupling coefficients $g_{1c}$, $g_{2c}$, and $g_{12}$ are associated with $C_{1c}$, $C_{2c}$, and $C_{12}$, respectively. 

The intermediate transmon employs a DC SQUID instead of a fixed Josephson junction, the effective Josephson energy is\cite{Koch2007}
\begin{equation}
	E_{J_1}\varPhi_{c}\simeq E_{J_1,\Sigma}\cos\left(\frac{\pi\varPhi_{c}}{\varPhi_0}\right),
	\label{eq:eff_ej1}
\end{equation}
where $E_{J_1,\Sigma}$ is the sum of two Josephson junction energies, $\varPhi_{c}$ is the magnetic flux threading the SQUID loop, and $\varPhi_0$ is the flux quantum. The effective Josephson inductance result in the flux changes and tunes eigen-frequency of the intermediate transmon \cite{Zhang2024}. Related driven Kerr circuits show how the same two-photon driving and Kerr terms can generate long-lived structured bosonic states \cite{Beaulieu2025}. This flux dependence provides tunability without changing the transmon characteristic of the central circuit element. The auxiliary transmons are operated mainly in low-excitation section, so the weak an-harmonicities are incorporated into the effective level structure.
Projecting Eq.(1) onto the low-energy subspace and retaining the dominant exchange terms \cite{Blais2021}
\begin{equation}
	\begin{split}
		\hat{H}(t) &= -\Delta_F \hat{F}^{\dagger}\hat{F} + \frac{K_{eff}}{2}\hat{F}^{\dagger 2}\hat{F}^2
		+ \frac{G_{eff}}{2}\left(\hat{F}^{\dagger 2}+\hat{F}^2\right) \\
		&\quad + \Delta_1(t)\hat{T}_1^{\dagger}\hat{T}_1 + \Delta_2\hat{T}_2^{\dagger}\hat{T}_2 \\
		&\quad + J_{F_1}\left(\hat{F}^{\dagger}\hat{T}_1 + \hat{T}_1^{\dagger}\hat{F}\right) + J_{12}\left(\hat{T}_1^{\dagger}\hat{T}_2 + \hat{T}_2^{\dagger}\hat{T}_1\right)
	\end{split}
	\label{eq:rotwave_ham}
\end{equation}
The first line describes the driven nonlinear target fluxonium, and the remaining terms describe the two auxiliary transmons and their nearest-neighbor exchange couplings. Operator $\hat{F}$ is annihilate excitation in the target fluxonium. The quantity $\Delta_F$ is the $F$ mode detuning, $K_{eff}$ is the effective Kerr nonlinearity of the fluxonium, and $G_{eff}$ is the effective two-photon driving. $\hat{T}_1$ and $\hat{T}_2$ is annihilate excitations in the intermediate transmon and dissipative transmon, respectively. The quantity $\Delta_1$ is the $T_1$ mode detuning, and $\Delta_2$ is the $T_2$ mode detuning. Eq.(3) defines the minimal chain model used in the baseline open-system calculations.  
The interference between the direct and intermediate-transmon-mediated contributions an additional term can be restored explicitly.
\begin{equation}
	\hat{H}^{dir} = J_{F_2}\left(\hat{F}^{\dagger}\hat{T}_2+\hat{T}_2^{\dagger}\hat{F}\right).
	\label{eq:H_dir}
\end{equation}
The direct $F-T_2$ coupling is set to zero in this model.

Flux tuning of the intermediate transmon through the time-dependent detuning
\begin{equation}
	\Delta_1(t) = \omega_1\varPhi_{c}(t) - \omega_r,
	\label{eq:delta1_t}
\end{equation}
where $\omega_1\Phi_c(t)$ is the flux-dependent frequency of the intermediate transmon, frequency $\omega_1$ can be changed by the SQUID flux. $\omega_r$ is reference frequency. When the intermediate transmon is far detuned from the two outer modes, its occupation is suppressed and the dissipative transmon has only a weak influence on the reduced fluxonium dynamics. Bringing the intermediate-transmon frequency into a region of appreciable hybridization enhances exchange with the auxiliary subsystem, allowing the dissipative transmon coupling with environment, and acting indirectly on the target mode. The additional dissipation result from a controlled spectrum rather than a permanent increasing in the intrinsic fluxonium loss.

The single-excitation response under a weak alternating current(AC) flux modulation is shown in Figs.1(b)–1(d). The flux is applied to the SQUID loop can be written as
\begin{equation}
	\varPhi_{c}(t) = \varPhi_{dc} + \varPhi_{p}\cos\left(2\pi f_{p}t+\phi_{p}\right).
	\label{eq:flux_modulation}
\end{equation}
where $\varPhi_{dc}$ is the static flux bias and $\varPhi_{p}$, $f_{p}$, and $\phi_{p}$ are the modulation amplitude, frequency, and initial phase, respectively. We choose the initial state $\left|100\right\rangle$, with the three occupation numbers refer to the target fluxonium, intermediate transmon, and dissipative transmon. 
For any basis state $|abc\rangle$, the corresponding population is
\begin{equation}
	P_{abc}(t,f_p)=\langle abc|\rho(t,f_p)|abc\rangle.
	\label{eq:pop_def}
\end{equation}
 Figs.1(b)-1(d) use $P_{100}$, $P_{010}$, and $P_{001}$, respectively.
Far away from the exchange resonance, the initial occupation remains predominantly in the target fluxonium, while the occupations of the two auxiliary modes are suppressed by detunings. As the modulation frequency approaches to resonance $f_{p}^{res}$, $P_{100}$ develops a pronounced depletion region, accompanied by oscillatory features in $P_{010}$ and $P_{001}$. Process of the intermediate-transmon transient population exchange is recorded, whereas the dissipative-transmon population indicates that occupation reaches to a more strongly damped mode. The different ripple widths and decay envelopes arise from the combined effects of detuning, coherent exchange, and mode-dependent loss. 
Without energy relaxation, the evolution is confined to the single-excitation manifold, the sum of the three populations remain a constant. Once dissipation is included, part of the occupation is transferred to the joint ground state and released into the environment. A small instantaneous value of $P_{001}$ has reached to the dissipative transmon and decayed the transferred occupation. The frequency scanning provides information about the exchange resonance, the transient involvement of the intermediate transmon, and the characteristic of decay timescale of the dissipative transmon.
Since the interaction associated with $C_{12}$ is weaker than the two nearest-neighbor couplings, its leading influence can be absorbed into small corrections to the effective frequencies and coupling strengths. 
For parameters of the normalized three-mode FTT model, all rates and detunings are expressed by normalized angular-frequency units as shown Table I.
\begin{table}[htbp]
	\centering
	\caption{parameters of the normalized three-mode FTT model.}
	\label{tab:ftt-parameters}
	\begin{tabular}{lcc}
		\hline
		Parameter & Symbol & Value \\
		\hline
		Target-mode detuning & $\Delta_F$ & $0.50$ \\
		Transfer-mode detuning (off) & $\Delta_{T_1,off}$ & $20.0$ \\
		Transfer-mode detuning (on) & $\Delta_{T_1,on}$ & $5.0$ \\
		Terminal-mode detuning & $\Delta_{T_2}$ & $5.0$ \\
		Fluxonium Kerr nonlinearity & $K_{eff}$ & $-0.10$ \\
		Two-photon drive amplitude & $G_{eff}$ & $0.70$ \\
		$F-T_1$ exchange coupling & $J_{F_1}$ & $4.40$ \\
		Direct $F-T_2$ coupling & $J_{F_2}$ & $0.0$ \\
        $T_1-T_2$ exchange coupling & $J_{12}$ & $4.50$ \\
		Fluxonium decay rate & $\kappa_F$ & $0.05$ \\
		$T_1$ decay rate & $\kappa_1$ & $0.05$ \\
		$T_2$ external decay rate & $\kappa_{2,ext}$ & $40.0$ \\
		$T_2$ internal decay rate & $\kappa_{2,int}$ & $10.0$ \\
		$T_2$ total decay rate & $\kappa_2$ & $50.0$ \\
		Thermal occupations & $N_j$ & $0.0$ \\
		Pure-dephasing rates & $\gamma_{\phi,j}$ & $0.0$ \\
		\hline
	\end{tabular}
\end{table}

The complete open-system evolution is governed by the Lindblad master equation
\begin{equation}
	\begin{split}
		\frac{d\hat{\rho}}{dt} &= -i\left[\hat{H}(t),\hat{\rho}\right]
+\sum_{j=F,T_1,T_2}\kappa_j\left(N_j+1\right)\hat{\mathcal{D}}\left[\hat a_j\right]\hat{\rho} \\
		&\quad +\sum_{j=F,T_1,T_2}\kappa_j N_j\hat{\mathcal{D}}\left[\hat{a}_j^{\dagger}\right]\hat{\rho}
	\end{split}
	\label{eq:lindblad}
\end{equation}
with
\begin{equation}
	\hat{\mathcal{D}}\left[\hat{O}\right]\hat{\rho} = \hat{O}\hat{\rho}\hat{O}^{\dagger} - \frac{1}{2}\left\{\hat{O}^{\dagger}\hat{O},\hat{\rho}\right\},
	\label{eq:lindblad_superop}
\end{equation}
 where $\hat a_j$ is its annihilation operator. The parameters $\kappa_j$ and $N_j$ denote the energy-decay rate and effective thermal occupation of mode $j$, respectively. The target fluxonium has a comparatively weak intrinsic loss, the intermediate transmon has a finite decay, and the dissipative transmon is assigned the largest decay rate. This asymmetric loss structure allows occupation enter into the auxiliary subsystem and then release efficiently while reduce persistent disturbance of the target fluxonium when the intermediate transmon is spectrally detuned, following the same separation of coherent storage and tunable exchange used in recent fluxonium-coupler studies \cite{Moskalenko2022,Zhang2024}.

In the following calculations,  we vary the detuning $\Delta_1(t)$ of the intermediate transmon through a periodic flux modulate, we need modify the spectral connection between the target fluxonium and the auxiliary dissipative subsystem. Without real-time measurement, the  parameters are not adjusted according to the instantaneous state. Within each parameter interval, the three-mode system evolves autonomously under the Hamiltonian and Lindblad master equation, after occupation enter into the dissipative transmon is released through coupling with the environment.

\section{STATE EVOLUTION AND RECOVERY UNDER PERIODIC FLUX MODULATION}
The classic circuit-quantum electrodynamics(QED) method needs continuously monitor, where measurement back-action and measurement records become a part of the evolution \cite{Blais2021}. In this paper, by solving three-mode Lindblad master equation and trace the time evolution over one closed cycle, we can examine whether the detuning tuning can connect steady states with different occupations. The intermediate-transmon detuning $\Delta_1(t)$, the system parameters are fixed. We focus on the mean fluxonium occupation and the transient responses of the two auxiliary modes. The three-mode model introduced in Sec.II shows that the frequency of the intermediate transmon controls the hybridization between the target fluxonium and the dissipative auxiliary subsystem.
In order to investigate how this tunable connection influences the target mode, we change the intermediate-transmon waveform  $\Delta_1(t)$ by using a periodic flux modulation. Since the waveform is specified before the evolution and without modify according to the instantaneous state, the modulation is open loop and involves neither real-time measurement nor measurement-based feedback. The corresponding auxiliary pathway leads to the formation of a low-occupation state and the reconstruction of the initial high-occupation state, respectively. Transient occupation of the two auxiliary transmons, irreversible energy relaxation, and the finite response time of the coupled system can leave a residual difference between the initial and final states. We examine the occupations of the three modes and then use the number-state and phase-space distributions to determine how the target fluxonium to finish change and recovery.

 Isolating the state of the target fluxonium, the two auxiliary transmons are traced out, all target-mode quantities including occupation, Fock-state distribution, and Wigner distribution can be calculated by the reduced density matrix (we denote the density matrix of the three-mode is $\rho(t)$.) 
\begin{equation}
	\hat{\rho}_F(t) = Tr_{T_1,T_2}\hat{\rho}(t)
	\label{eq:rho_reduced}
\end{equation}
 The mean occupations of the three modes are
\begin{equation}
	\langle \hat{n}_j(t) \rangle = Tr\left[\hat{\rho}(t)\hat{a}_j^{\dagger}\hat{a}_j\right], j=F,T_1,T_2.
	\label{eq:mean_occupation}
\end{equation}
The Fock-state distribution contains the diagonal density-matrix elements in the number basis, the mean target occupation in Fig.2(b) follows these probabilities
\begin{equation}
	\langle n_F(t)\rangle=\sum_n n\,P_F(n,t).
	\label{eq:occupation_from_fock}
\end{equation}
The target-fluxonium Fock-state probability is 
\begin{equation}
	P_F(n,t) = \langle n \rvert \rho_F(t) \lvert n \rangle.
	\label{eq:p_fock}
\end{equation}
Eq.(\ref{eq:p_fock}) resolves the quantity plotted in Fig.2(b) into its individual number-state contributions.  In the initial high state, the probability is distributed over several Fock states and decreases gradually with increasing $n$. Once the auxiliary dissipation becomes effective, the probability becomes strongly concentrated at $n=0$, while the weights of the higher Fock states are suppressed by several orders of magnitude. The reduction in the mean occupation corresponds to a broad redistribution of the Fock-state population. After the intermediate transmon returns to the detuned region, the higher-number components are re-established. Over the displayed range, the recovered distribution nearly overlaps the initial high-state distribution, both at small $n$ and in the decaying tail. The increase in the mean occupation is accompanied by the recovery of the original multilevel number distribution.
\begin{figure}[t]
	\centering
	\includegraphics[width=0.8\linewidth]{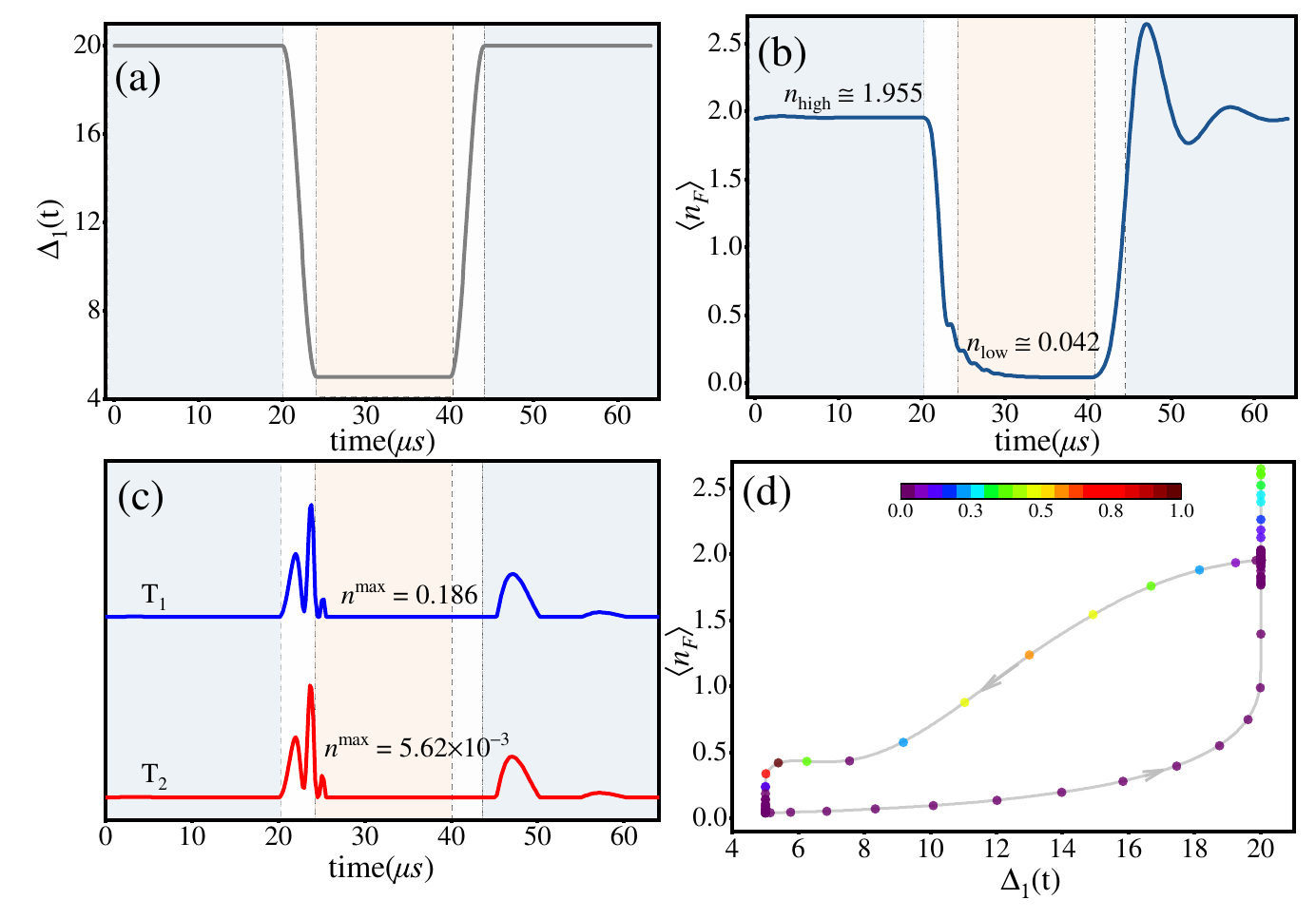}
	\caption{\label{fig:ftt_cycle_dynamics}Mode-occupation evolution during one period of flux modulation. (a) Time-dependent detuning $\Delta_1(t)$ of the intermediate transmon when the system moves between the far-detuned and hybridized parameter regions. (b) Mean occupation $\langle n_F\rangle$ of the target fluxonium. The marked values indicate the  high- and low-occupation levels during the modulation cycle. (c) Instantaneous occupations of the intermediate and dissipative transmons when the auxiliary dissipative pathway becomes effective. (d) Direct trajectory in the $(\Delta_1(t),\langle n_F\rangle)$ plane. The two branches correspond to the decreasing- and increasing-detuning parts of the modulation, and the arrows indicate the direction of evolution.
	}
\end{figure}

Fig.2(a) shows the intermediate-transmon detuning over one modulation period and allows the three-mode to establish the modified spectral hybridization over a nonzero response time. Initially, $\Delta_{T_1}$ at a far-off resonant region, suppressing its hybridization with the two outer modes. and then enter into hybridization region, remains in a finite duration, and finally returns to initial value. The target-fluxonium occupation changes with this modulation as shown in Fig.2(b). The system initially occupies a high-state region with $\bar{n}_{F}^{H} \simeq 1.955$. When the intermediate transmon enters into the hybridized region, exchange between the target mode and the auxiliary subsystem becomes stronger. Consequently, $\langle n_F \rangle$ decreases rapidly and approaches to the low-state value $\bar{n}_{F}^{L} \simeq 0.042$. When the intermediate-transmon detuning returns to the off-resonant region, its hybridization with the outer modes is reduced and the additional dissipation ceases. The two-photon driving and Kerr nonlinearity restore fluxonium occupation toward to initial high value. The transient overshoot and subsequent damped oscillations show that depletion and recovery are governed by different physical mechanism. The decrease in occupation is driven by auxiliary dissipation, whereas the recovery depends on the nonlinear dynamics of the fluxonium. In Fig.2(c), the maximum intermediate-transmon occupation is approximate to $\bar{n}_{T_1}^{max} \simeq 0.186$, with the peak occur when the detuning changes and the hybridization becomes appreciable. The maximum dissipative transmon instantaneous occupation is $\bar{n}_{T_2}^{max} \simeq 5.62\times10^{-3}$. An effective dissipative output does not require a large population number accumulate. The transient peaks in $T_1$ indicate that the intermediate transmon participates as an actual dynamical degree of freedom. Fig.2(d) plots the target occupation against the control detuning. The decreasing and increasing parts of the waveform yield different values of $\langle n_F \rangle$ at the same $\Delta_1$. 
Because the modulation occurs at a finite rate, the curve should not be identified directly with equilibrium hysteresis, nor does a single period establish a strict limit cycle. Its final point lies close to the initial point, showing that the mean target occupation is largely recovered, although occupation alone can't determine whether the complete fluxonium density matrix has returned. Importantly, it is not merely the instantaneous minimum value during a transient depletion. Held at the pathway-on detuning, the system remains close to this low-occupation plateau. The activation of the auxiliary pathway drives the system toward a low-occupation steady state become a new parameter configuration, rather than produce a short-lived depletion of the target-mode occupation. We extract states from  high- and low-occupation plateaus and prolong the evolution with the fixed parameters. For the high-occupation configuration, the mean fluxonium occupation remains close to its initial stationary value. Under the pathway-on configuration, the low-occupation state continues to converge toward a stable value. The two plateaus correspond to long-time states associated with two distinct fixed-parameter configurations. These fixed-parameter evolution  evidence that the two plateaus are genuine high- and low-occupation steady states. The mean occupation measures the overall energy scale of the target mode. Distinct density matrices have identical or similar occupations, thus  $\langle n_F \rangle$ return to a high value.

Fig.3(a)-(c) compares Fock-state and Wigner distributions through the numerical evaluation of Eq.(\ref{eq:p_fock}) at the initial high state, the low state, and the recovered high state. The displaced-parity representation is the standard phase-space definition of the Wigner quasi-probability.
\begin{equation}
	W_F(\alpha,t)=\frac{2}{\pi}Tr\big[\rho_F(t) D(\alpha) (-1)^{\hat{F}^{\dagger}\hat{F}} D^{\dagger}(\alpha)\big],
	\label{eq:wigner_fluxonium}
\end{equation}
where $D(\alpha)$ is the displacement operator. The initial high state exhibits two separated regions along the $\mathrm{Re}\,\alpha_F$ direction. This two-lobe structure results from the combined action of the two-photon driving and Kerr nonlinearity, the present distribution is not sufficient to identify the state as a pure coherent cat state \cite{Reglade2024}.

\begin{figure}[htbp]
	\centering
	\includegraphics[width=0.9\linewidth]{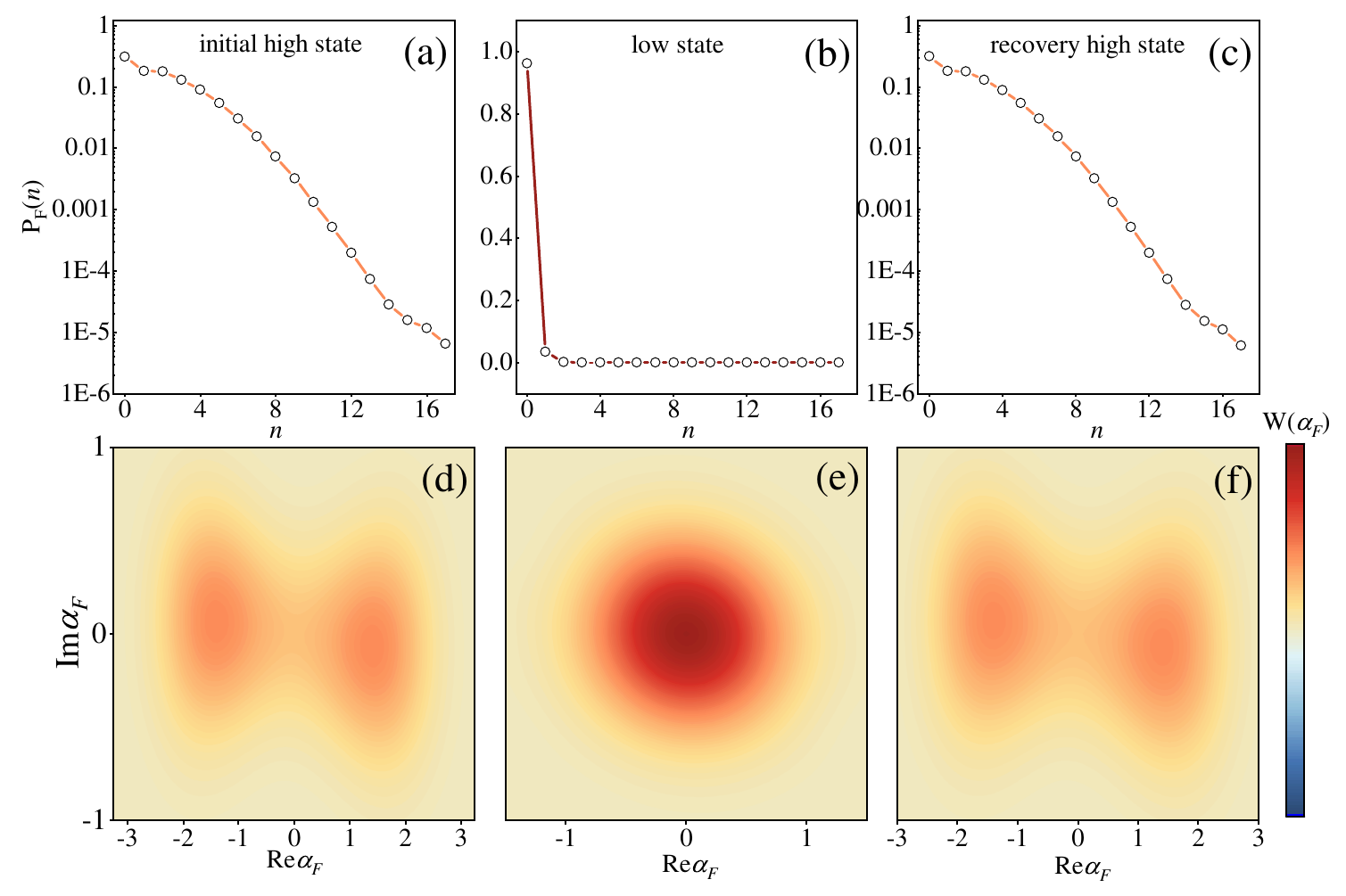}
	\caption{\label{fig:ftt_quantum_state_revised}Target-fluxonium state distributions at three representative stages of the modulation cycle. (a)-(c) Fock-state probabilities $P_F(n)$ for the initial high-occupation state, the low-occupation state, and the recovered high-occupation state, respectively. (d)-(f) Corresponding Wigner distributions $W_F(\alpha)$. When the auxiliary dissipative pathway is effective, the higher-number components are suppressed and the phase-space distribution contracts toward to the origin. After the intermediate transmon becomes detuned again, the higher-number components and the two-lobe phase-space structure reappear and approach their initial distributions.}
\end{figure}

When the auxiliary dissipation becomes effective, the Wigner distribution contracts toward the phase-space origin and the separated lobes disappear. This contraction agrees with the concentration of the Fock-state probability near $n=0$, showing that the low state is characterized by both a smaller mean occupation and a reduced phase-space extent. Once the intermediate transmon is detuned again, the Wigner distribution expands and recovers a two-lobe profile. The peak positions, spatial widths, and overall shape of the recovered distribution are close to those of the initial high state. The Fock-state and Wigner distributions describe the same evolution: auxiliary dissipation moves the target fluxonium from an extended high-occupation state into a low-occupation region, the reduction of the dissipative connection allows the nonlinear mode return to its initial state.

The near overlap of the displayed distributions is not exact equality of the two density matrices. The Fock-state probabilities do not contain all off-diagonal coherence, and Wigner distributions appear similar at finite resolution can retain small differences. The results support substantial recovery of the mean occupation, number-state distribution, and phase-space structure.
Such finite-accuracy recovery is consistent with open-system dynamics. Information released to environment during the modulation can't be retrieved by restoring the control parameter.

Figs.2-3 connect the control parameter with the change in the quantum state. Flux tuning of the intermediate transmon modify the spectral connection between the target fluxonium and the auxiliary subsystem, lowering the target occupation. When the intermediate transmon becomes detuned again, the additional dissipation is reduced and the target mode returns to high-occupation region. The transient auxiliary populations, parameter-space curve, Fock-state probabilities, and Wigner distributions reveal complementary aspects of mode exchange, population release, and state recovery. 

\section{POPULATION TRANSFER AND ENVIRONMENTAL RELEASE}
In this section, we examine the respective roles of  intermediate transmon, dissipative transmon, and asymmetric mode losses, and uses parameter variations and control calculations to identify the structural origin of the observed response. Sec.III show that coherent coupling can temporarily redistribute occupation among the three modes and allow part of it to return at a later time. An irreversible contribution arises only when occupation transfer to the dissipative transmon subsequently release to environment. 
We take the time derivative of the target-mode occupation. For the minimal $F$-$T_1$-$T_2$ chain,
\begin{equation}
	\frac{d}{dt}\langle n_F\rangle
	=
	P(t)-\kappa_F\left[\langle n_F\rangle-N_F\right]-I_{F_1}(t).
	\label{eq:target_rate}
\end{equation}
Here $P(t)=2G_{eff}\mathrm{Im}\langle \hat{F}^{\dagger 2}\rangle$ is the occupation change produced by the two-photon driving. The other two terms describe intrinsic loss from the target mode and transfer from $F$ to $T_1$, respectively. The transfer current follows from the exchange term in the Hamiltonian. Together with the current between the two auxiliary modes,
\begin{equation}
	\begin{aligned}
	I_{F_1}(t)=-2J_{F_1}\mathrm{Im}\langle
	\hat{F}^{\dagger}\hat{T}_1\rangle,I_{12}(t)
=-2J_{12}\mathrm{Im}\langle\hat{T}_1^{\dagger}\hat{T}_2\rangle.
	\end{aligned}
	\label{eq:mode_currents}
\end{equation}
The positive $I_{F_1}$ means transfer out of the target fluxonium, the positive $I_{12}$ means transfer from $T_1$ to $T_2$. The current release to environment through $T_2$, the damped-mode input-output relation satisfy \cite{Blais2021}
\begin{equation}
	\begin{gathered}
		\frac{d\langle n_{T_1}\rangle}{dt}=I_{F_1}(t)-I_{12}(t)-\kappa_{T_1}\left[\langle n_{T_1}\rangle-N_{T_1}\right],\\[3pt]
		\frac{d\langle n_{T_2}\rangle}{dt}=I_{12}(t)-r(t),
		r(t)=\kappa_{T_2}\left[\langle n_{T_2}\rangle-N_{T_2}\right].
	\end{gathered}
	\label{eq:auxiliary_continuity}
\end{equation}
If $J_{12}=0$, we have $I_{12}=0$, and then we requires a strongly reduced $r(t)$. $r(t)$ is the instantaneous environment release as shown in Fig. 4(a) and Fig. 4(d). Integrating this final rate gives out the cumulative released occupation
\begin{equation}
	Q(t)=\int_0^t r(t')\,dt',
	\label{eq:cumulative_release}
\end{equation}
Eq.(18) is used in Fig. 4(b) and Fig. 5(b), which describe target-mode loss appears as $I_{F_1}$, then as $I_{12}$, and finally as irreversible release $r(t)$.
 
\begin{figure}[htbp]
	\centering
	\includegraphics[width=0.8\linewidth]{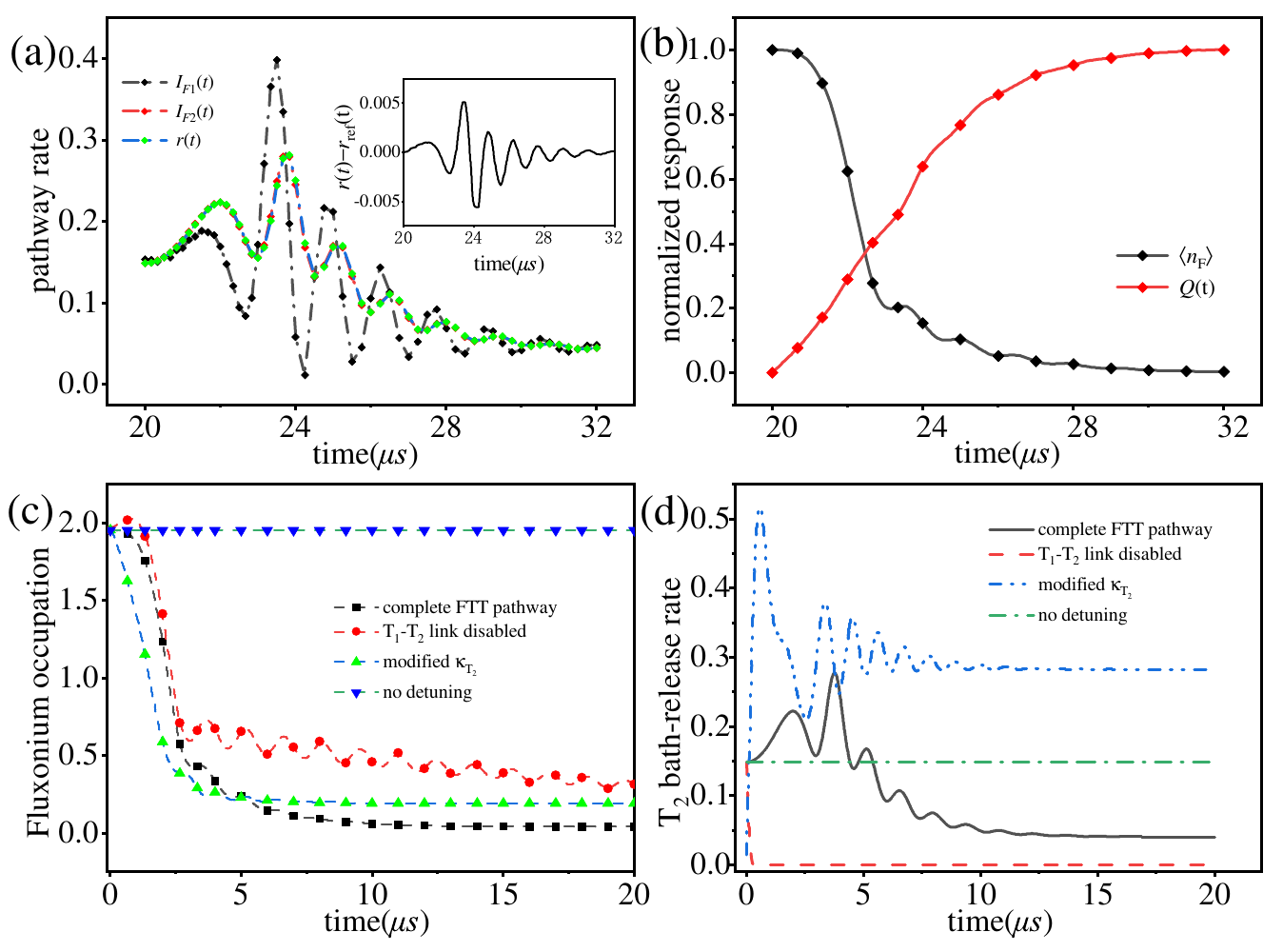}
	\caption{\label{fig:ftt_mechanism_counterfactual}Occupation transfer, environmental release, and comparison calculations for the auxiliary dissipative pathway.
(a) The successive stages of the auxiliary dissipative pathway. Transfer from the target fluxonium to the intermediate transmon to the dissipative transmon and then release to environment. The inset shows the modulation-induced change in the dissipative-transmon release. (b) Normalized occupation in the target fluxonium and normalized cumulative release to the environment through the dissipative transmon. The two quantities are separate. (c) Target-fluxonium occupation for three-mode pathway, disabled $T_1-T_2$ coupling, modified dissipative-transmon decay rate $\kappa_{T_2}$, and without detuning modulation. (d) Dissipative-transmon release rate under the same four conditions as in panel (c).}
\end{figure}
Fig.4(a) denotes instantaneous transfer rates from the target fluxonium to the intermediate transmon, from the intermediate transmon to the dissipative transmon, and from the dissipative transmon to its environment. The three contributions do not reach to their maximum simultaneously. The temporal offsets reflect the finite response of the coupled system as occupation is redistributed through the auxiliary subsystem before release it to the environment. The inset isolates the modulation-induced change in the dissipative-transmon release. Fig.4(b) compares the normalized occupation in the target fluxonium with the normalized cumulative release through the dissipative transmon. The two quantities do not satisfy a strict conservation relation. It nevertheless associates the formation of the low-occupation state with continued release to environment. Without detuning modulation, the target fluxonium close to its initial high-occupation state. The dissipative transmon coupling to environment, the auxiliary subsystem spectrally weakly connected to the target mode. A lossy mode is insufficient to account for the observed depletion, tunability of the intermediate transmon provides the spectral selectivity required to activate the auxiliary pathway. Changing the dissipative-transmon decay rate $\kappa_{T_2}$ modifies both the target-mode response and the environmental release. The comparison establishes that the observed evolution is sensitive to the dissipative-transmon loss, a single modify value is not sufficient to determine an optimal decay rate or a complete dependence on $\kappa_{T_2}$. Figs.4(c)-(d) test whether this response requires the complete auxiliary pathway. When the coupling between the two transmons is removed, the intermediate transmon can still exchange occupation with the target fluxonium, but the transferred occupation cannot efficiently reach to the principal dissipative outlet. The target mode shows partial depletion, while the environmental release through the dissipative transmon is strongly reduced.

We assume the two loss contributions in terms of Eq.\,(\ref{eq:target_rate}) 
\begin{equation}
	R(t)=\kappa_F\left[\langle n_F\rangle-N_F\right]+I_{F_1}(t),
	\label{eq:target_removal}
\end{equation}
which reduces the target-mode equation to
\begin{equation}
	\frac{d}{dt}\langle n_F\rangle=P(t)-R(t).
	\label{eq:target_balance}
\end{equation}
Eq.(\ref{eq:target_balance}) is the rate balance plotted in Fig.5(a). The sign of $P(t)-R(t)$ must equal to the sign of the measured change in target occupation. During pathway activation, $R(t)>P(t)$ and the measured occupation decreases. When the intermediate transmon is detuned again, removal is reduced, $P(t)>R(t)$, and the occupation recovers.  The overshoot shows that recovery is a new driven-loss balance, not a time-reversed depletion process.
The target occupation falls whenever removal exceeds driven-induced injection.
Integrating the same balance from the beginning of the cycle 
\begin{equation}
	\langle n_F(t)\rangle-\langle n_F(0)\rangle
	=
	\int_0^t P(t')\,dt'-\int_0^t R(t')\,dt'.
	\label{eq:integrated_target_balance}
\end{equation}
This relation predicts that the cumulative injection and removal curves as shown in Fig.5(b). It also shows why a recovered target occupation does not require the environmental release to reverse. Flux tuning of the intermediate transmon determines when the target mode connect to the auxiliary subsystem, while coupling to the dissipative transmon access to a stronger environmental loss channel. The low-occupation state results from the combined action of spectral matching, coherent exchange, and environmental release.

\begin{figure}[htbp]
	\centering
	\includegraphics[width=0.8\linewidth]{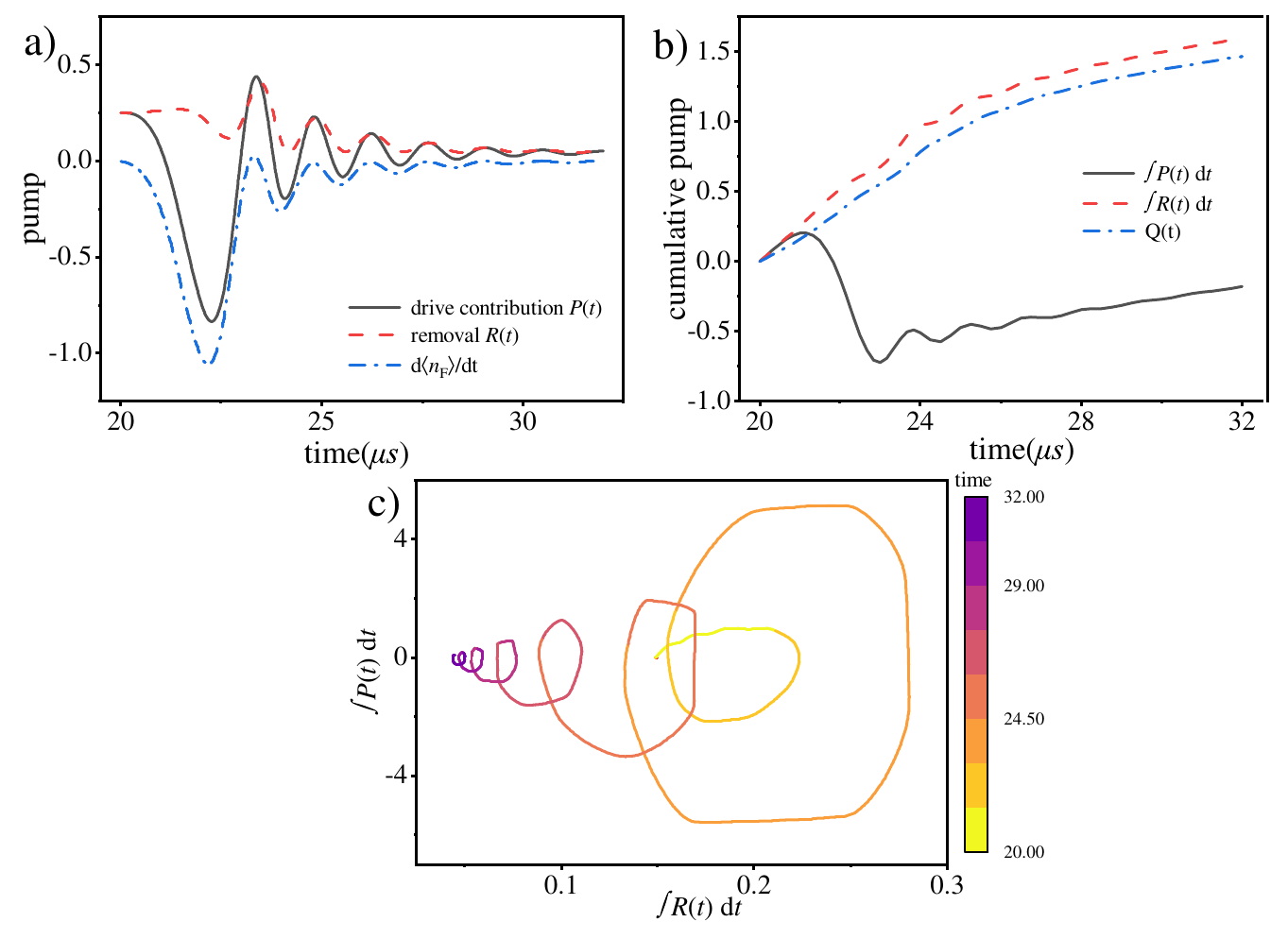}
\caption{\label{fig:ftt_full_cycle_balance}Target-mode occupation balance and cumulative flow during activation of the auxiliary dissipative pathway. (a) Two-photon occupation injection, total target-mode removal, and the numerical time derivative $d\langle n_F\rangle/dt$. Their relative magnitudes determine the decrease and recovery of the target-fluxonium occupation during the modulation.
(b) Cumulative contributions from the two-photon driving, target-mode removal, and environmental release through the dissipative transmon. (c) Directed trajectory formed by the cumulative injection and removal. The color scale denotes time and distinguishes different stages of the evolution.}
\end{figure}
Fig.5 examines the same process from the target-mode occupation balance. Fig.5(a) compares the two-photon injection, total target-mode removal, and the numerical derivative $d\langle n_F\rangle/dt$. During pathway activation, removal exceeds driven-induced injection and the target occupation decreases. Near the modulation boundaries, the oscillatory structure reflects the coexistence of coherent inter-mode exchange and dissipative removal. After the intermediate transmon returns to the far-detuned region, the additional removal through the auxiliary pathway is reduced and the two-photon driving becomes dominant for a finite interval. The target occupation consequently rises again, with an overshoot and damped oscillations before a new balance is reached. Depletion and recovery occur under different hybridization conditions and different rate balances. Fig.5(b) shows the time-integrated contributions. During the modulation cycle, continued driving replenishes the target mode and the cumulative injection approaches to the cumulative removal. The cumulative release through the dissipative transmon continues to increase. The recovered high-occupation state  once the auxiliary connection reduce, continued driving rebuilds the target occupation under the restored weak-dissipation condition. Fig.5(c) provides a two-dimensional representation of the cumulative injection and removal as a trajectory. The color scale denotes time and distinguishes different stages of the trajectory. The activation and recovery parts do not retrace the same path in reverse, consistent with the finite response time and irreversible environmental release. Figs. 4 and 5 separate the transport and balance aspects of the same evolution. When the auxiliary pathway is active, spectral matching provided by the intermediate transmon allows occupation to enter into the auxiliary subsystem and subsequently leave through the dissipative transmon. Target-mode removal then exceeds driven-induced injection and the fluxonium approaches to the low-occupation state. When the intermediate transmon becomes detuned, the additional removal is suppressed and the driving rebuilds the target occupation. The recovered high state results from renewed driven-induced occupation.
\section{NOISE ROBUSTNESS AND STATE RECOVERY}
The auxiliary dissipation is an essential part of the depletion-recovery process. We treat pure dephasing, finite-temperature excitation, and stochastic detuning fluctuations as additional perturbations and examine how they affect the evolution of the target fluxonium. The dynamics are governed by
\begin{align}
	\frac{d\hat{\rho}}{dt}
	&=-i[\hat H(t),\hat{\rho}]
	+\sum_{j}\kappa_j(N_j+1)\hat{\mathcal D}[\hat a]\hat{\rho}
	\nonumber\\
	&\quad
	+\sum_{j}\kappa_jN_j\hat{\mathcal D}[\hat a_j^\dagger]\hat{\rho}
	+\sum_{j}\gamma_{\phi,j}\hat{\mathcal D}[\hat a_j^\dagger\hat a]\hat{\rho} ,
	\label{eq:master_noise}
\end{align}
where $j=F,T_1,T_2$. $\gamma_{\phi,j}$ is the pure-dephasing rate, and $N_j$ denotes the effective thermal occupation\cite{Ithier2005}. This occupation is $N_j(T)=1/[e^{\hbar\omega_j/(k_{B}T)}-1]$, which provides the direct connection between temperature and the thermal terms in the master equation\cite{Nava2022}. When the engineering dissipation is introduced, these additional perturbations are excluded.
In order to distinguish a noise-induced displacement of the high-occupation state, we introduce trace-norm distinguishability measures, which are standard quantum information diagnostics for comparing quantum states and their correlations \cite{Groisman2005}.
\begin{align}
	D_H=\frac12\left\|\rho_{n}-\rho_{r}
	\right\|_1,
	D_{rec}=\frac12\left\|\rho_{f}-\rho_{i}
	\right\|_1 .
	\label{eq:distances_def}
\end{align}
Here, $D_H$ measures the displacement of the noisy high-occupation state from its reference counterpart, whereas $D_{rec}$ measures how closely the final state returns to the initial high-occupation state under the same noise condition. 

\begin{figure}[htbp]
	\centering
	\includegraphics[width=0.8\linewidth]{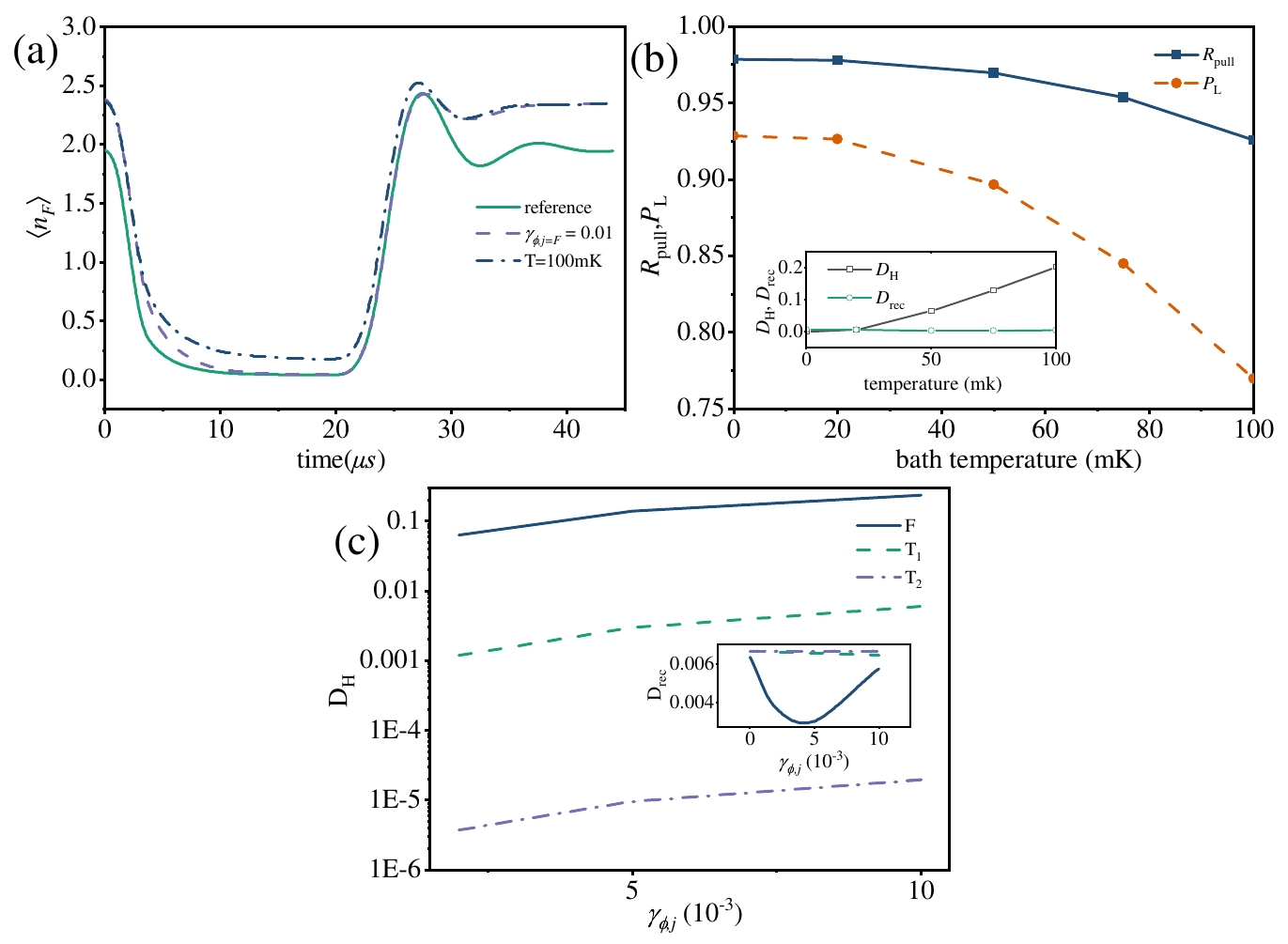}
\caption{\label{fig:noise_dephasing_thermal}Effects of pure dephasing and finite temperature on the evolution and recovery of the target fluxonium.
		(a) Mean target-mode occupation for the reference, target-fluxonium pure dephasing with $\gamma_{\phi,F}=0.01$,
		and a bath temperature of $T=100mK$. (b) Recovery metric $R_{pull}$ and low-state mean occupation $\bar{n}_{F}^L$ as functions of temperature, the inset shows the corresponding $D_H$ and $D_{rec}$. (c) $D_H$ when pure dephasing is applied separately to the target fluxonium, intermediate transmon, and dissipative transmon, the inset shows the corresponding $D_{rec}$.
	}
\end{figure}

As shown in Fig.~\ref{fig:noise_dephasing_thermal}(a), $\bar{n}_{F}^H$ and $\bar{n}_{F}^L$ are the mean target occupations on the high- and low-occupation plateaus of the same cycle. The characteristic of high-low-high evolution remains visible in pure dephasing and finite temperature, although the states are shifted at different stages of the cycle. Increasing the bath temperature raises up the residual occupation of the low state, leading to a larger $\bar{n}_{F}^L$ and a gradual reduction in recovery. In Fig.6(b), the strength of the high-to-low occupation conversion is characterized separately by $R_{pull}=1-\bar{n}_{F}^L/\bar{n}_{F}^H$, which quantifies how efficiently the target mode is pulled into the low-occupation region, it is instead described by $D_{rec}$. $D_H$ increases with temperature while $D_{rec}$ remains comparatively small. The final state remain close to the initial state under the same noisy environment. Fig.\ref{fig:noise_dephasing_thermal}(c) shows that dephasing applied directly to the target fluxonium produces the largest change in $D_H$, whereas auxiliary transmon $D_{rec}$ remain small.

\begin{figure}[htbp]
	\centering
	\includegraphics[width=0.8\linewidth]{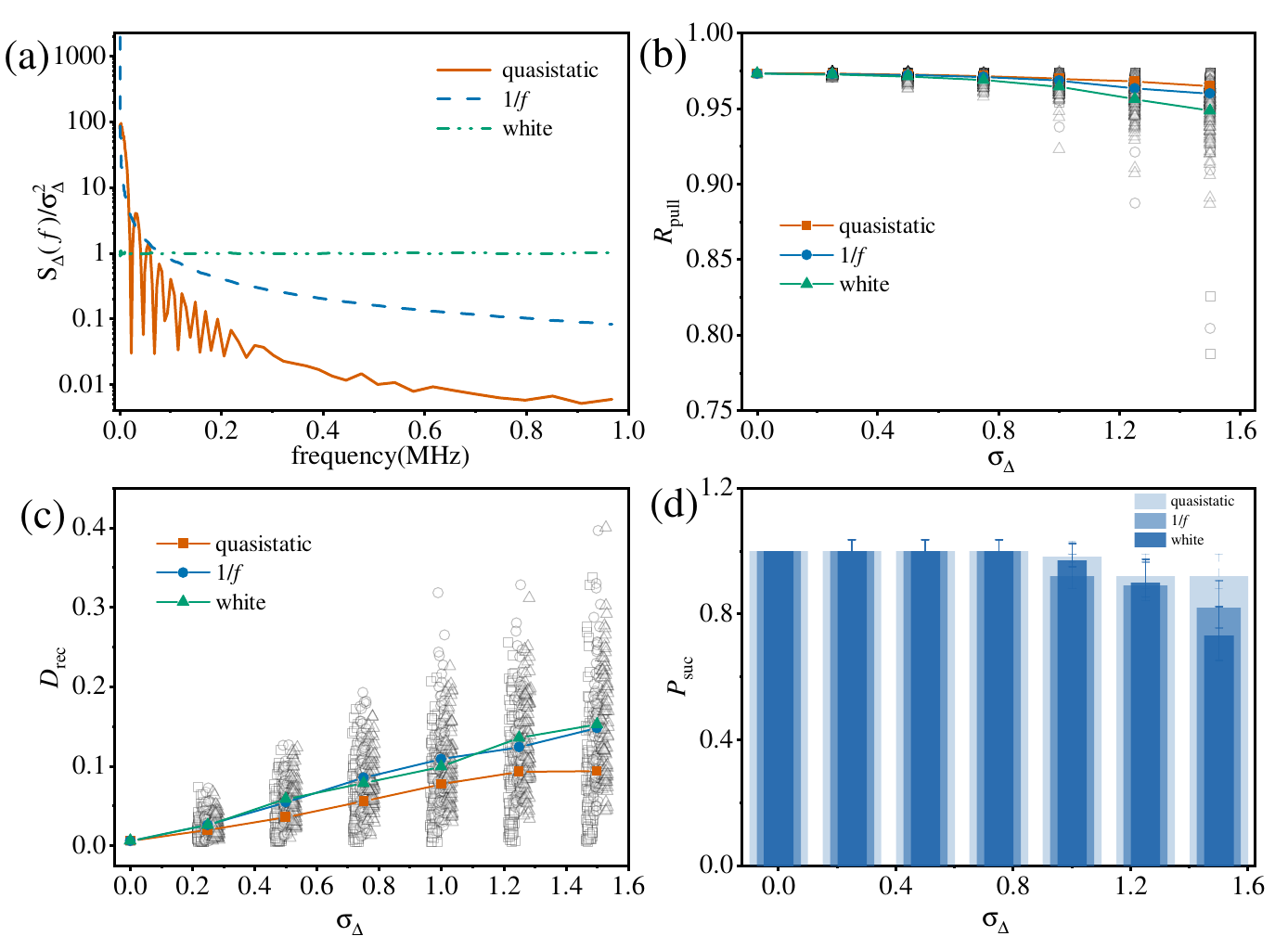}
\caption{\label{fig:detuning_noise_models} Effects of effective detuning noise on target‑state recovery.
		(a) Normalized power spectral densities of quasi-static, $1/f$, and white detuning noise.
		(b) Recovery metric $R_{pull}$ as a function of the noise amplitude $\sigma_\Delta$.
		(c) Recovery distance $D_{ rec}$ as a function of $\sigma_\Delta$.
		(d)Success probability $P_{suc}$ for the three noise models.
		For each noise model, six nonzero noise strengths are sampled with $100$ independent realizations per strength. Light open symbols denote the individual realizations, and each filled symbol is the mean of the corresponding $100$ values.
	}
\end{figure}

A quasi-static offset mainly moves the operating point for an entire cycle, while dynamic noise continually changes the instantaneous spectral matching. For each noise realization, $R_{pull}$ is evaluated by using its high and low plateaus. The filled points in Fig.~\ref{fig:detuning_noise_models} are averages over the corresponding noise records, whereas the open points show individual values.

As shown in Fig.7, this parameterization follows standard flux-noise descriptions and noise-spectroscopy measurements in superconducting circuits \cite{Dai2021,Karamlou2023,Ezzell2023}. 
Random fluctuations in the flux-control channel are included as a time-dependent detuning error $\delta\Delta_1(t)$ added to the prescribed control detuning $\Delta_1(t)$. For quasi-static noise, $\delta\Delta_1(t)=\delta\Delta_0$ over one modulation cycle. The dynamic models are specified by $S_\Delta(f)=A/f$ for $1/f$ noise and $S_\Delta(f)=S_0$ for white noise. Their common strength parameter is the rms amplitude $\sigma_\Delta$, defined over the simulated frequency band by $\sigma_\Delta^2=\int_{f_{ l}}^{f_{h}}S_\Delta(f)\,df$. For each of the three noise models, we consider six nonzero values of $\sigma_\Delta$, and generate $100$ independent realizations at each value. The statistical analysis contains $3\times6\times100=1800$ noisy trajectories. For every realization, $R_{\rm pull}$ is evaluated using its high and low plateaus. At each nonzero noise strength, the filled point in Fig.~\ref{fig:detuning_noise_models} is the arithmetic mean of the corresponding 100 values, whereas the open points show the individual realizations. As the detuning noise increases, the spread among stochastic realizations becomes broader, accompanied by a decrease in $R_{pull}$ and an increase in $D_{rec}$. The effect of detuning noise depends not only on its overall amplitude but also on its temporal structure. A quasi-static offset primarily shifts the operating point over an cycle, whereas time-dependent fluctuations continuously perturb the instantaneous spectral matching between the modes. The influence of frequency noise depends on operating point. A detuning fluctuation apply to the intermediate transmon modify the spectral condition of the auxiliary pathway and reaches to the target fluxonium through inter-mode coupling. By contrast, an equivalent fluctuation apply directly to the target mode changes its local frequency without intermediate propagation step.

Near a given operating point, this distinction can be understood in terms of the effective self-energy generated by the auxiliary subsystem. To make the connection explicit, we first consider the local linear response and denote the frequency-domain amplitudes of the three modes by $F$, $T_1$, and $T_2$. The two auxiliary-mode equations are
\begin{align}
	\left(\omega-\Delta_1+\frac{i\kappa_1}{2}\right)T_1
	&=J_{F_1}F+J_{12}T_2, \nonumber\\
	\left(\omega-\Delta_2+\frac{i\kappa_2}{2}\right)T_2
	&=J_{12}T_1 .
	\label{eq:auxiliary_response}
\end{align}
The second relation gives out the response of $T_2$ to $T_1$. Eliminating $T_2$ and give out the auxiliary-network contribution seen by the target mode,
\begin{equation}
	\Sigma(\omega)=
	\frac{J_{F_1}^{2}}
	{\omega-\Delta_1+\frac{i\kappa_1}{2}
		-\dfrac{J_{12}^{2}}
		{\omega-\Delta_2+\frac{i\kappa_2}{2}}},
	\label{eq:self_energy_aux}
\end{equation}
The symbol $\Sigma(\omega)$ denotes contribution. Substitution it into the target-mode equation then gives out local response,
\begin{equation}
	\chi(\omega)=
	\frac{1}
	{\omega-\Delta_F+\frac{i\kappa_F}{2}-\Sigma(\omega)}.
	\label{eq:susceptibility_target}
\end{equation}
For a small fluctuation around the operating point, we have $\delta\chi(\omega)\simeq[\partial\chi(\omega)/\partial\Delta_1]\delta\Delta_1$. A fluctuation on $T_1$ first changes the auxiliary response and then affects the target mode. A smaller $|\partial\chi/\partial\Delta_1|$ predicts weaker local transfer of detuning noise. The full state evolution is obtained from the three-mode Lindblad equation, standard circuit-response, open-system elimination, and reservoir-engineering methods \cite{Clerk2010,ReiterSorensen2012,MetelmannClerk2015}.

\begin{figure}[htbp]
	\centering
	\includegraphics[width=1\linewidth]{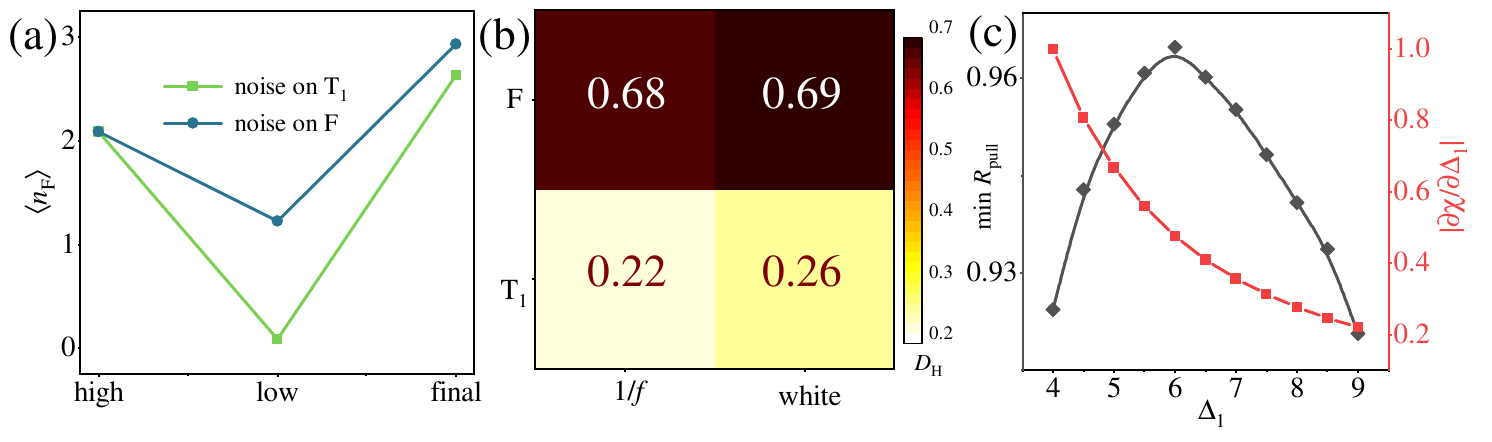}
\caption{\label{fig:noise_injection_site}Dependence of target‑state robustness on the noise‑injection site and the operating detuning of the intermediate transmon. (a) Mean target‑fluxonium occupation at the initial high‑occupation, low‑occupation, and final stages when the same type of frequency noise is applied separately to the intermediate transmon $T_1$ and the target fluxonium $F$. (b) State deviations for $1/f$ and white noise applied to $T_1$ and $F$, respectively. (c) Worst‑case recovery metric $R$ and local detuning sensitivity $|\partial\chi/\partial\Delta_1|$ as functions of the operating detuning $\Delta_1$. The horizontal dashed line marks the recovery threshold $R=0.95$, the shaded region indicates the feasible operating range, and the vertical dashed line marks the selected point $\Delta_1=7$.}
\end{figure}

Fig.\ref{fig:noise_injection_site}(a) shows that when frequency noise through the intermediate transmon, the target fluxonium can still reach to a low-occupation state. Applying the same type of fluctuation directly to the target mode raises up the low-state occupation substantially. The same mode dependence for both $1/f$ and white noise in Fig.\ref{fig:noise_injection_site}(b), where direct perturbation of the target fluxonium produces a larger state deviation. The intermediate transmon does more than tune the auxiliary dissipative pathway, which makes the coupling between control fluctuations and the target mode indirect. This noise transfer is dependent on operating point. Fig.\ref{fig:noise_injection_site}(c) compares the recovery metric with $|\partial\chi/\partial\Delta_1|$ over the same detuning range. High recovery within a finite interval, while the local sensitivity varies with detuning. The operating point balance efficient state recovery against sensitivity to control fluctuations. $\Delta_1=7$ lies in the region satisfy the recovery constraint while retain relatively low detuning sensitivity.

Residual detuning fluctuations remain particularly for quasistatic and low-frequency components.
To account for the frequency dependence of calibration and control-line filtering, we introduce an effective transfer function $H(f)$,
\begin{align}
	\sigma_{eff}^2=\int S_{out}(f)df=\int |H(f)|^2S_{in}(f)df .
	\label{eq:noise_transfer_function}
\end{align}
Here, $S_{in}(f)$ and $S_{out}(f)$ is the one-sided detuning-noise spectra before and after processing, respectively. $\sigma_{eff}$ is the remaining rms detuning over the frequency band. The separation of slow parameter estimation from frequency-selective suppression is motivated by real-time Hamiltonian estimation and filter-function treatments of dynamical decoupling \cite{Hangleiter2024}. Combining this result with the linear response gives out the local estimate
\begin{equation}
	\left\langle|\delta\chi|^2\right\rangle
	\simeq
	\left|\frac{\partial\chi}{\partial\Delta_1}\right|^2
	\sigma_{eff}^2.
	\label{eq:noise_to_response}
\end{equation}
The two relations predict that calibration reduces the state deviation by lowering the weighted noise amplitude by the sensitive operating point. The quantitative values of $R_{pull}$ and $D_{rec}$ are evaluated from the complete three-mode evolution.

\begin{figure}[htbp]
	\centering
	\includegraphics[width=1\linewidth]{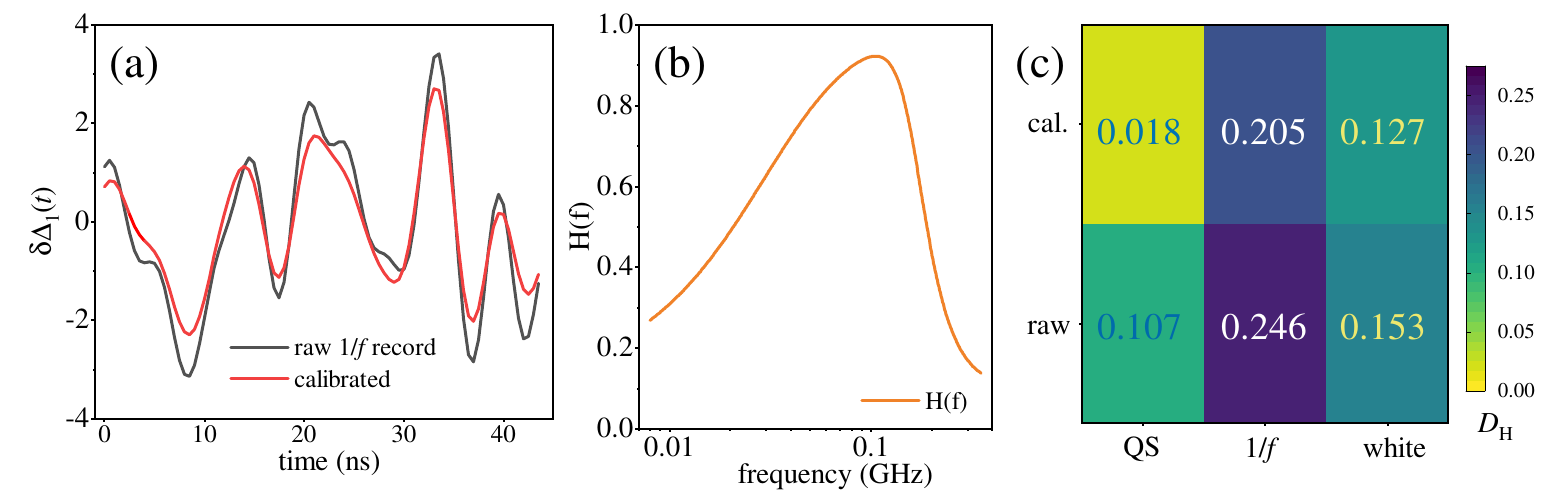}
	\caption{\label{fig:noise_calibration_filter}Calibration and frequency‑dependent process of detuning noise in the intermediate‑transmon control channel.
		(a) Representative $1/f$ detuning‑noise record before and after calibration.
		(b) Effective transfer function $H(f)$ describe the frequency‑dependent weight of the control noise.
		(c) Comparison of the state deviation before and after process for quasi-static, $1/f$, and white noise.
	}
\end{figure}

Figs.\ref{fig:noise_calibration_filter}(a)-(b) show that different frequency components with different weight of the control chain. In Fig.~\ref{fig:noise_calibration_filter}(c), the state deviation associate with quasi-static noise decreases from approximate $0.107$ to $0.018$, the corresponding values for $1/f$ and white noise decrease from $0.246$ to $0.205$ and from $0.153$ to $0.127$, respectively.
The larger improvement for quasi-static noise as operating‑point offset, which can be corrected effectively by calibration.
Dynamic components of $1/f$ and white noise partially reduced.

 Figs.\ref{fig:noise_dephasing_thermal}-\ref{fig:noise_calibration_filter} describe how additional noise enters into the three‑mode system. Pure dephasing and finite temperature shift the states during the modulation cycle, whereas stochastic detuning perturbs the spectral matching of the auxiliary pathway. The extent to these fluctuations reach to the target fluxonium depends on both the injection site and the operate detuning, while the frequency response of the control chain determines the residual detuning.

\begin{table*}[htbp]
\centering
\caption{Comparison of representative three-mode cycle.}
\label{tab:protocol-comparison}
\renewcommand{\arraystretch}{0.50}
\resizebox{\textwidth}{!}{%
\begin{tabular}{ccccc}
\hline
Year & Scheme & Control & Dynamics & Result \\
\hline
2013-15\cite{Shankar2013,Leghtas2015} & Reservoir Engineering  & Loss & Steady state & Stationary target \\
2017\cite{Fitzpatrick2017} & Circuit QED Lattice  & Drive $+$ Loss & Random switching & Dim-bright phases \\
2020\cite{Grimm2020,Lescanne2020} & Kerr Cat  & Kerr / Two-photon Loss & Steady manifold & Cat protection \\
2023\cite{Chen2023} & Duffing Transition  & Kerr $+$ Loss & One-way relaxation & Critical slowing \\
2025\cite{Beaulieu2025} & Two-photon Kerr  & Two-photon Drive $+$ Loss & Dissipative transition & Hysteresis \\
2026 &  Our work & Flux $F-T_1-T_2$ & Closed high-low-high & State depletion and recovery \\
\hline
\end{tabular}%
}
\end{table*}

The characteristic depletion and recovery of the target state identifiable over finite ranges of pure dephasing, thermal excitation, and control‑detuning fluctuations. We can reduce the exposure of the target state to control‑line noise by coupling the control to the target fluxonium indirectly through the intermediate transmon, and by combining this structure with an appropriate operate point and frequency‑dependent calibration. 

\section{conclusion}

We investigates a three-mode open quantum system consist of a target fluxonium, a flux-tunable intermediate transmon, and a dissipative transmon. Based on full Lindblad master-equation, our numerical simulations demonstrate that periodic flux modulation drives the target fluxonium from a high-occupation non-equilibrium state to a suppressed low-occupation regime, followed by recovery toward to the initial high-occupation configuration. Both Fock-state populations and Wigner quasi-probability distributions confirm that recovery is established consistently in both number-space and phase-space characteristics.

Quantitative analysis of mode-resolved occupation transfer and environmental release clarifies the physical roles of the auxiliary subsystems: the intermediate transmon enables frequency-selective hybridization with the target mode, the dissipative transmon channels transfer excitations into the thermal bath. Systematic comparison simulations verify that the full depletion–recovery dynamics relies on flux tunability, inter-transmon coupling, and dissipative decay of the auxiliary mode. Furthermore, the essential dynamical features persist under finite energy relaxation, pure dephasing, elevated temperature, and stochastic detuning noise.

The achieved high-occupation states exhibit non-equilibrium characteristics. Our framework offers a feasible route toward controllable auxiliary dissipation and robust non-equilibrium state manipulation, provide a valuable reference for future investigations of superconducting quantum systems.

\begin{acknowledgments}
		This work was funded by the State Key Laboratory of Quantum Optics Technologies and Devices, Shanxi University, Shanxi, China (Grants No.KF202503); Zhejiang Key Laboratory of Quantum State Control and Optical Field Manipulation, Hangzhou Dianzi
        University (KYZ074326001).
	\end{acknowledgments}
    
\section{DATA AVAILABILITY}
The data that support the findings of this article are not publicly available. The data are available from the authors upon reasonable request.

\bibliography{ref}
\end{document}